\documentclass[twocolumn]{aastex61}
\usepackage{amsmath}

\newcommand\kms{km\,s$^{-1}$}

\received{}
\revised{}
\accepted{}
\submitjournal{ApJ}

\shorttitle{Detailed 3D Reconstruction of SNR N132D}
\shortauthors{Law et al.}
\begin{document}

\title{Three-Dimensional Kinematic Reconstruction of the Optically-Emitting, High-Velocity, Oxygen-Rich Ejecta of Supernova Remnant N132D}

\correspondingauthor{Charles J.\ Law}
\email{charles.law@cfa.harvard.edu}

\author[0000-0003-1413-1776]{Charles J.\ Law}
\affiliation{Center for Astrophysics $|$ Harvard \& Smithsonian, 60 Garden St., Cambridge, MA 02138, USA}

\author{Dan Milisavljevic}
\affil{Department of Physics and Astronomy, Purdue University, 525 Northwestern Avenue, West Lafayette, IN 47907, USA}

\author{Daniel J.\ Patnaude}
\affil{Center for Astrophysics $|$ Harvard \& Smithsonian, 60 Garden St., Cambridge, MA 02138, USA}

\author{Paul P.\ Plucinsky}
\affil{Center for Astrophysics $|$ Harvard \& Smithsonian, 60 Garden St., Cambridge, MA 02138, USA}

\author{Michael D.\ Gladders}
\affil{Department of Astronomy \& Astrophysics, The University of Chicago, 5640 South Ellis Avenue, Chicago, IL 60637, USA}

\author{Judy Schmidt}
\affil{Astrophysics Source Code Library, Michigan Technological University, 1400 Townsend Drive, Houghton, MI 49931, USA}

\author{Niharika Sravan}
\affil{Department of Physics and Astronomy, Purdue University, 525 Northwestern Avenue, West Lafayette, IN 47907, USA}

\author{John Banovetz}
\affil{Department of Physics and Astronomy, Purdue University, 525 Northwestern Avenue, West Lafayette, IN 47907, USA}

\author{Hidetoshi Sano}
\affil{Institute for Advanced Research, Nagoya University, Furo-cho, Chikusa-ku, Nagoya 464-8601, Japan}
\affil{Department of Physics, Nagoya University, Furo-cho, Chikusa-ku, Nagoya 464-8601, Japan}

\author{Jordan M.\ McGraw}
\affil{Research Computing, ITaP, Purdue University, West Lafayette, IN 47907, USA}

\author{George Takahashi}
\affil{Research Computing, ITaP, Purdue University, West Lafayette, IN 47907, USA}

\author{Salvatore Orlando}
\affil{INAF---Osservatorio Astronomico di Palermo ``G.S. Vaiana", Piazza del Parlamento 1, I-90134 Palermo, Italy}



\begin{abstract}

We present a three-dimensional kinematic reconstruction of the optically-emitting, oxygen-rich ejecta of supernova remnant N132D in the Large Magellanic Cloud. Data were obtained with the 6.5~m Magellan telescope in combination with the IMACS+GISMO instrument and survey [O~III]~$\lambda\lambda$4959,~5007 line emission in a ${\sim}$3$^{\prime}~\times$~3$^{\prime}$ region centered on N132D. The spatial and spectral resolution of our data enable detailed examination of the optical ejecta structure. The majority of N132D's optically bright oxygen ejecta are arranged in a torus-like geometry tilted approximately 28$^{\circ}$ with respect to the plane of the sky. The torus has a radius of 4.4~pc~($D_{\rm LMC}$/50~kpc), exhibits a blue-shifted radial velocity asymmetry of $-3000$~to~$+2300$~km~s$^{-1}$, and has a conspicuous break in its circumference. Assuming homologous expansion from the geometric center of O-rich filaments, the average expansion velocity of 1745~km~s$^{-1}$ translates to an age since explosion of 2450~$\pm$~195~yr. A faint, spatially-separated ``runaway knot" (RK) with total space velocity of 3650~km\,s$^{-1}$ is nearly perpendicular to the torus plane and coincident with X-ray emission that is substantially enhanced in Si relative to the LMC and N132D's bulk ejecta. These kinematic and chemical signatures suggest that the RK may have had its origin deep within the progenitor star. Overall, the main shell morphology and high-velocity, Si-enriched components of N132D have remarkable similarity with that of Cassiopeia A, which was the result of a Type~IIb supernova explosion. Our results underscore the need for further observations and simulations that can robustly reconcile whether the observed morphology is dominated by explosion dynamics or shaped by interaction with the environment. 
\end{abstract}




\section{Introduction} \label{sec:intro}

Young, nearby supernova remnants (SNRs) offer rare opportunities to investigate kinematic and chemical details of supernova (SN) explosions at fine scales that are impossible to achieve in unresolved extragalactic events \citep{MF17}. Their morphology and elemental distribution provide unique insight into where and how asymmetry is introduced during core collapse \citep{LF18}, which is presently a key area of investigation in increasingly sophisticated 2D and 3D simulations \citep{MJ15,Janka16,Kuroda17,OC18,Burrows19}.  Expanding stellar ejecta that encode valuable information about the explosion dynamics, nucleosynthetic yields, and mixing of the progenitor star can be resolved, measured, and tracked. The progenitor star system's transitions through evolutionary stages leading up to core collapse and associated mass loss can also be explored by studying the SNR's interaction with its surrounding environment. 

O-rich SNRs are particularly well-suited for studies of core-collapse SN explosion dynamics and offer powerful tests of simulations that are now able to evolve from SN to the remnant phase \citep{Orlando15,Orlando16,Ferrand19}. This, in part, is because they are often associated with progenitor stars that were largely stripped of their hydrogen envelopes. SNe of this variety (Type IIb, Ib, Ic; see \citealt{Gal-Yam17} for details) retain critical elements of their central explosion dynamics that are otherwise disrupted in H-rich explosions \citep{Milisavljevic10}.  Only a handful of O-rich remnants have so far been identified in our Galaxy and in nearby galaxies. Cassiopeia A (Cas~A), which is regarded as the prototypical O-rich SNR \citep{Kirshner77, Chevalier78}, Puppis A \citep{Winkler85}, and G292$+$1.8 \citep{Goss79, Murdin79} are among the Galactic members. N132D \citep{Danziger76} and 0540$-$69.3 \citep{Mathewson80} are in the Large Magellanic Cloud, and 1E 0102.2$-$7219 (E0102; \citealt{Dopita81}), 0103$-$72.6 \citep{Park03}, and B0049-73.6 \citep{Hendrick05} are in the Small Magellanic Cloud. Several mixed-morphology SNRs, including CTB~1, HB~3, and W28 have also shown evidence of O-rich ejecta \citep{Lazendic06, Pannuti10, Pannuti17}. 

Located in the stellar bar of the LMC, N132D is the brightest X-ray and gamma-ray emitting SNR in that galaxy \citep{Clark82, Hwang93, Favata97, Borkowski07, HESS15, Ackermann16}. The remnant nature of N132D was first recognized by \citet{Westerlund66}, who noticed an association between a non-thermal radio source and bright [\ion{S}{2}] optical emission. The discovery of high-velocity [\ion{O}{3}] and [\ion{Ne}{3}] ejecta established the relative youth of N132D, indicating that the reverse shock has not yet reached the center of the remnant \citep{Danziger76, Lasker78}. Subsequent optical/UV spectra from the \textit{Hubble Space Telescope} (\textit{HST}) show strong emission from C- and Ne-burning elements but little emission from O-burning elements \citep[e.g.,][]{Morse96, France09}, which indicates that the remnant may be the debris of a ${\sim}25$\,M$_{\odot}$  progenitor star that experienced significant mass loss and exploded as a Type Ib SN \citep{Blair00}. 

High-resolution X-ray images reveal a horseshoe-shaped forward shock \citep[e.g.,][]{Borkowski07, Xiao08, Bamba18}, which has been modeled as a blast wave impacting an irregular cavity wall \citep{Chen03}, while spatially resolved X-ray spectroscopy reveals an asymmetric Fe-K distribution from an estimated $15\pm5$\,M$_{\odot}$ progenitor \citep{Sharda19_inprep}. This outer shell is also detected in hot dust emission in 24~$\mu$m \textit{Spitzer} maps \citep{Tappe06}, and submillimeter observations suggest a physical and kinematic association between the southern portion of N132D's outer shell and nearby, likely natal, CO molecular clouds \citep{Banas97, Sano17, Sano19_inprep}. Numerous luminous, shocked dense clouds with normal LMC ISM abundances \citep[e.g.,][]{Blair00, Dopita18} and velocity dispersions of ${\sim}$200~km~s$^{-1}$ are present throughout N132D. Prominent among these clouds are the outer western edge of the remnant and Lasker's bowl \citep{Lasker80}, which is located in the northern part of the remnant and exhibits the curved inner boundary of a classical bow shock morphology. Since these clouds do not show enhancements in [\ion{N}{2}], they were likely pre-existing ISM clouds over which the SNR blast wave recently passed \citep{Dopita18}.

\begin{figure*}[ht]
\centering

\includegraphics[width=\linewidth]{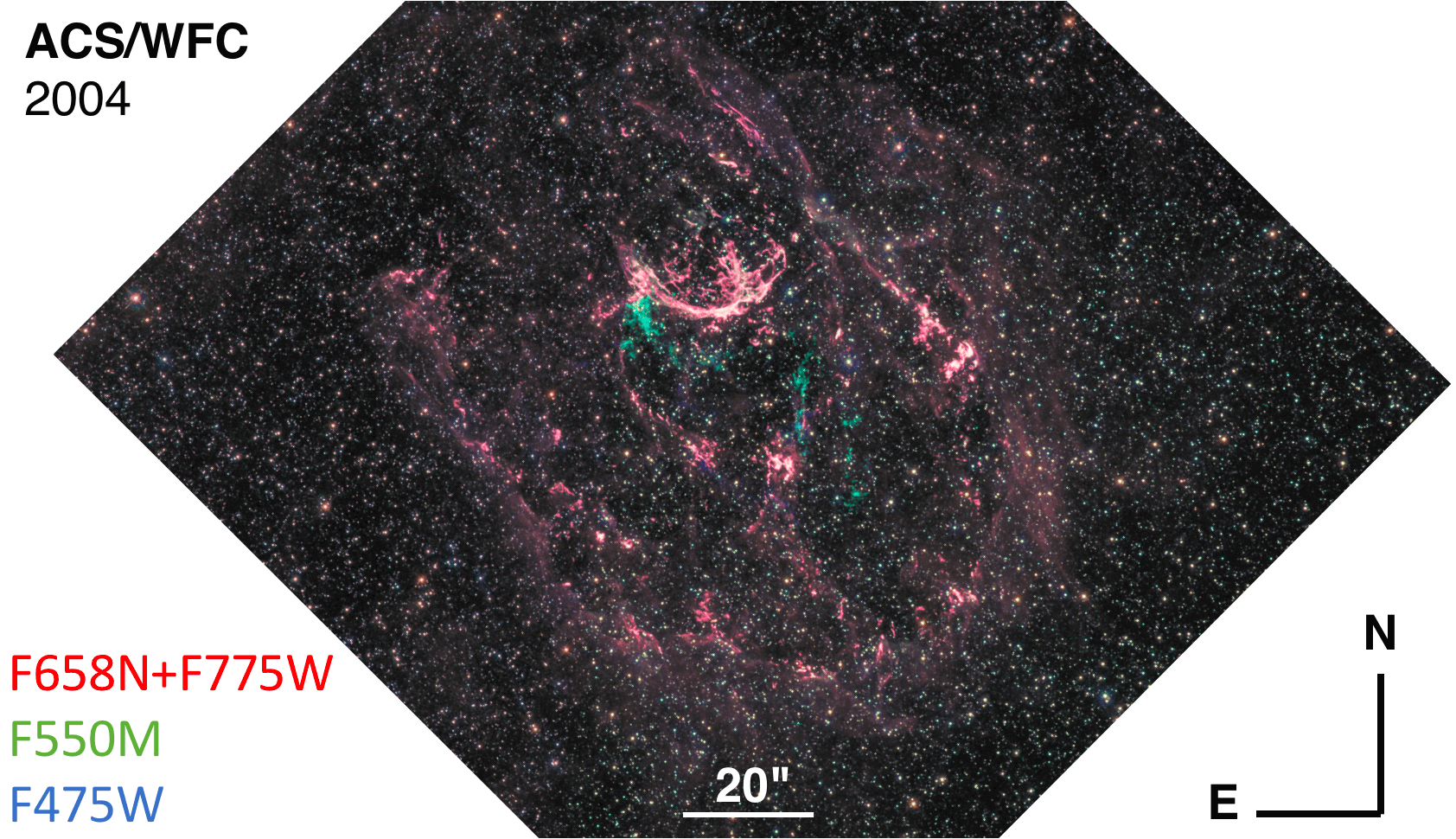}\\
\includegraphics[scale=0.35]{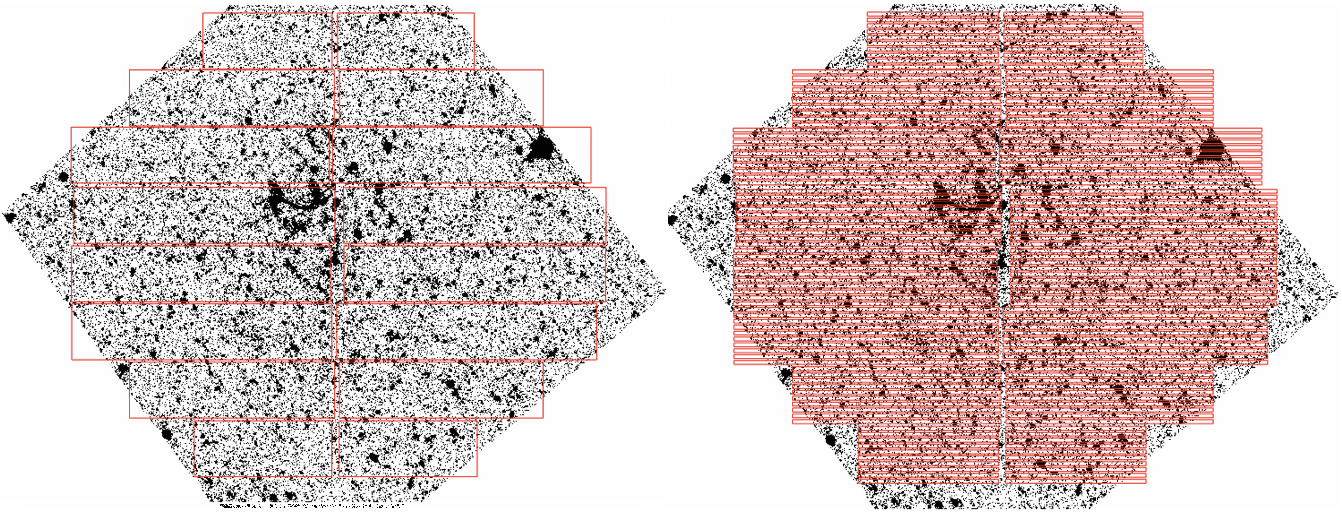}

\caption{\textit{Top}: \textit{HST} ACS composite image of N132D sensitive to oxygen and hydrogen emission using F475W, F550M, and F658N+F775W filters in blue, green, and red, respectively. Filamentary blue and green emission represents the location of O-rich ejecta, while diffuse red emission traces the horseshoe-shaped forward shock and ISM/CSM-related structures.\textit{Bottom left}: GISMO footprint on N132D. \textit{Bottom right}: Finding chart of all long-slit positions. Background image is the \textit{HST}/ACS F475W image. Data were retrieved from the Mikulski Archive for Space Telescopes and are associated with Proposal 12001 (PI J. Green).}
\label{fig:hstcomposite}
\end{figure*}

\begin{figure*}[ht]
\centerline{%
\includegraphics[width=\linewidth]{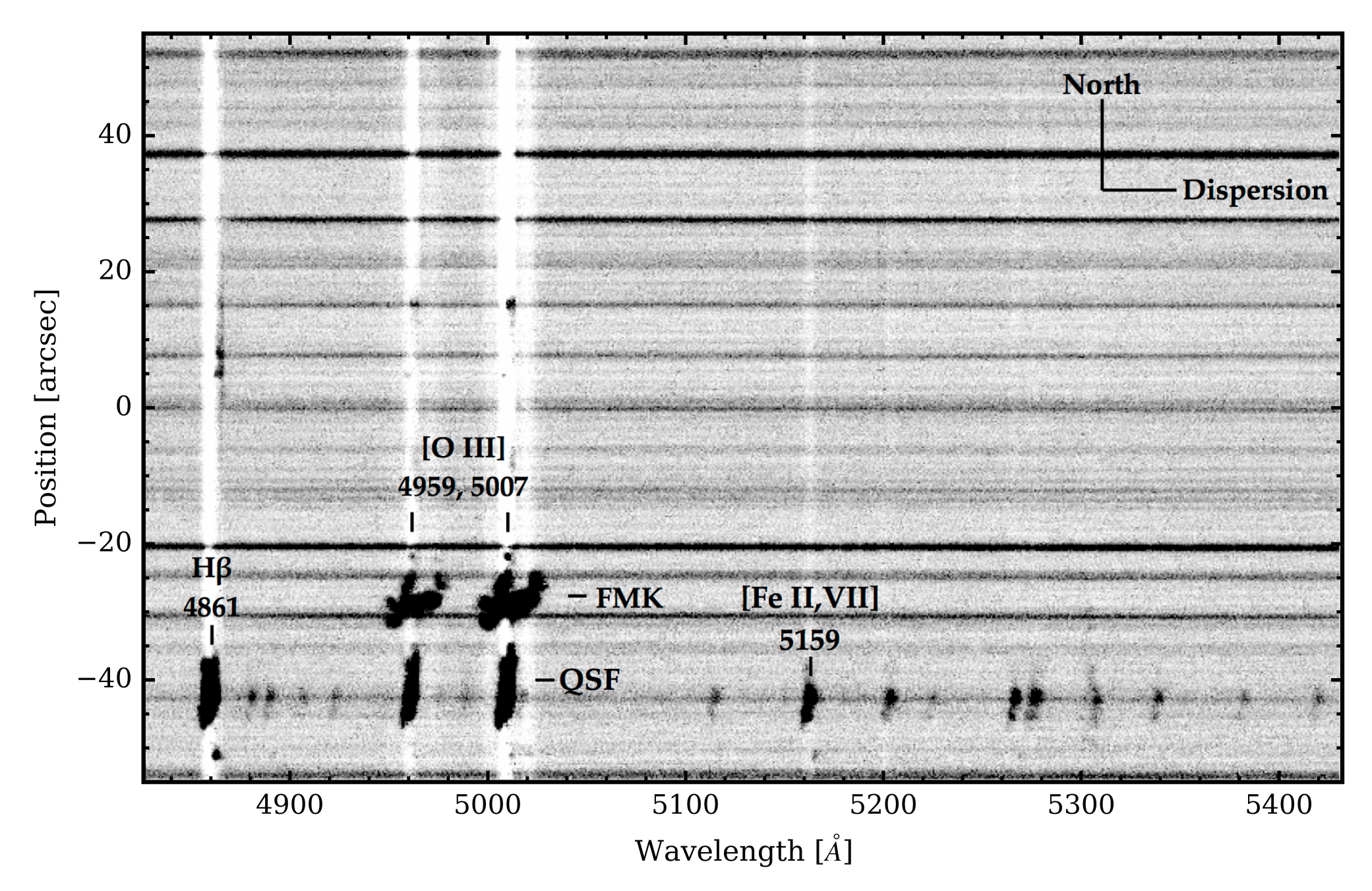}%
}%
\caption{Example of a fully-reduced and cleaned 2D spectrum of a slit position along the main shell from which 1D extractions were made. Horizontal lines represent continuum emission from stars contained within the slit. Fast moving knots (FMKs), the focus on this study, and ISM/CSM-related quasi-stationary flocculi (QSF) are labeled. Wavelength positions (\AA) for the brightest emission lines are marked.} 
\label{fig:chip}
\end{figure*}

The optical remnant is approximately 110$^{\prime\prime}$ in diameter and has three distinct emission components: (1) high-velocity filaments rich in oxygen and neon but lacking hydrogen, (2) low-to-moderate-velocity H-rich knots and filaments, consistent with clumpy mass loss by the progenitor star, and (3) diffuse emission from normal LMC abundance clouds at rest with respect to the local medium. Several Doppler reconstructions of N132D's O-rich optical material have been conducted over the last four decades. The first attempts revealed a thin, inclined ring \citep{Lasker80}, while later studies claimed that the O-rich knots instead form a thin shell \citep{Morse95}. A total velocity range of ${\sim}4500$~km~s$^{-1}$ was observed in the O-rich knots with blue-shifted material having higher velocities than the redshifted ejecta \citep[e.g.,][]{Morse95, Vogt11}. A survey of [\ion{O}{3}] $\lambda5007$ dynamics by \citet{Vogt11} revealed that the majority of O-rich ejecta form a ring ${\sim}12$~pc in diameter and are inclined at an angle of ${\sim}$25$^{\circ}$ to the line of sight. \citet{Vogt11} suggest that this ring of O-rich material is from ejecta in the equatorial plane of a bipolar explosion and that the observed morphology of N132D was heavily influenced by the pre-SN mass loss from the progenitor star. Furthermore, \citet{Vogt11} interpret the ``runaway knot'' (RK), which was first recognized by \citet{Morse95} as an outlying O-rich ejecta clump with higher-than-average space velocity, as possible evidence of a polar jet.

Unlike other young SNRs such as E0102 \citep{Finkelstein06} and Cas A \citep{Fesen01, Fesen06}, N132D lacks proper motion measurements of its ejecta. Hence, age estimates have been based on radial velocities that are translated into expansion velocities and the overall angular extent of the filamentary structures. From the dynamics of fast-moving O and Ne material, \citet{Danziger76} estimated a maximum age of 3440~yr and a probable age of ${\sim}1350$~yr. \citet{Lasker80} derived a similar age of ${\sim}1300$~yr, while subsequent determinations from \citet{Morse95} and \citet{Sutherland95} found ages of ${\sim}3150$~yr and ${\sim}$2350~yr, respectively. More recently, \citet{Vogt11} estimated an age of 2500~yr. The inconsistency in these age estimates is directly related to differing assumptions about the shape of the O-rich ejecta. 

To resolve this discrepancy in remnant age estimates that range between 1300--3440 yr and to explore N132D's morphology in hopes of gaining insight into the dynamics of the original supernova explosion, we undertook a detailed spectroscopic mapping of the entire N132D remnant with higher spatial resolution than previous efforts. We describe the observations and our data reduction methods in Sections \ref{sec:obs} and \ref{sec:data_redux}, respectively. In Section \ref{sec:reconstruction}, we discuss how these data were used to develop a 3D kinematic reconstruction of the optically-emitting oxygen ejecta and present the results of our reconstruction. In Section \ref{sec:discussion}, we discuss the kinematic properties of these ejecta in the context of SN explosion processes and the ejecta structure of other well-studied SNRs. We then summarize our findings in Section \ref{sec:conclusions}.

\begin{figure*}[ht]
\centerline{%
\includegraphics[width=\linewidth]{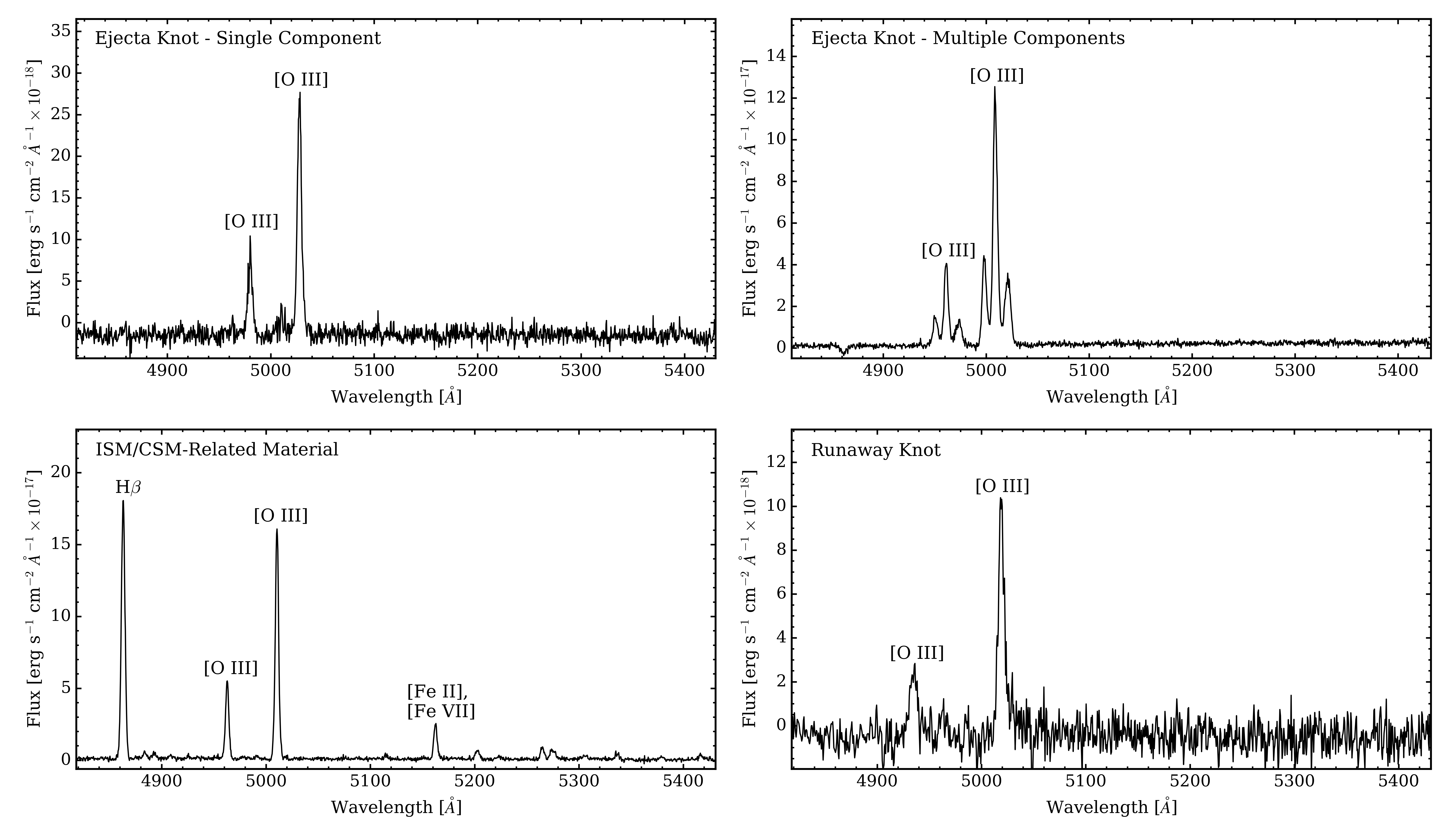}%
}%
\caption{Examples of 1D spectra of ejecta knots of N132D used in the Doppler reconstruction. The top left spectrum from an ejecta knot shows a single velocity component, while the top right shows a more complicated ejecta knot with three  separate velocity components. Bottom left shows ISM/CSM-related material, as indicated by numerous Fe lines and prominent H$\beta$ emission. Bottom right presents a spectrum from the runaway knot, identifying it as bona fide O-rich ejecta.} 
\label{fig:example_spectra}
\end{figure*}

\section{Observations} \label{sec:obs}

In Figure \ref{fig:hstcomposite}, we present a composite image of N132D from high spatial resolution \textit{HST} imaging taken with the Advanced Camera for Surveys (ACS) that maps the optical emission of N132D. These images, which were obtained on 2004 January 22, provide complete spatial coverage of the remnant \citep{Beasley04} and are associated with Proposal 12001 (PI J. Green). The ACS Wide Field Channel (WFC) F475W filter (SDSS $g$) covers the full range of Doppler shifted [\ion{O}{3}] $\lambda\lambda$4959, 5007 emission from high-velocity, O-rich ejecta. The F550M filter is sensitive to continuum emission. The F658N filter is sensitive to shocked ambient medium emitting strongly in H$\alpha$. The F775W filter (SDSS $i$) is sensitive to ambient medium emission lines of [Fe II], [Ca II], and [O~II] (see, e.g., \citealt{Dopita18}), with a small fraction of emission due to O-rich ejecta weakly emitting in [O~II] $\lambda\lambda$7319, 7330.

Low-dispersion long-slit optical spectra across the entire supernova remnant N132D were obtained on 31 December 2015 and 1 January 2016 with the Inamori Magellan Areal Camera and Spectrograph (IMACS; \citealt{Dressler11}) on the 6.5~m Baade Magellan telescope at Las Campanas Observatory. The f/4 camera was used with the Gladders Image-Slicing Multislit Option (GISMO)\footnote{\url{http://www.lco.cl/telescopes-information/magellan/instruments/imacs/gismo/gismoquickmanual.pdf}} in combination with a 600 lines~mm$^{-1}$ grating, the GISMO-3 filter ($\lambda_{\rm o} = 4981$\,\AA; $\Delta\lambda = 947$\,\AA) and a mask of long-slits oriented east-west. Slit widths were fixed at $1.6^{\prime \prime}$ and slit lengths ranged from $57^{\prime \prime}$ to $110^{\prime \prime}$, depending on slit location within the GISMO footprint. The resulting dispersion was 0.378~\AA\,per pixel with an effective wavelength window of approximately 4810--5430~\AA\ and a spectral resolution full-width-half-maximum (FWHM) of 5.5~\AA. The telescope was moved in the north-south direction in non-overlapping increments of 2.4$^{\prime \prime}$ in order to cover the majority of the remnant (67\%). A finding chart of the GISMO footprint along with all long-slit positions is shown in Figure \ref{fig:hstcomposite}. At each position, exposures of length $2\times800$~s were taken. Wavelength calibration He-Ne-Ar lamps were taken every two positions to track instrument flexure. The airmass was between 1.3--1.6 for all observations and the seeing was variable from 0.7$^{\prime \prime}$ to 1.0$^{\prime \prime}$. The spectrophotometric standard star LTT 3864 was observed as a flux calibration source.

\section{Data Reduction} \label{sec:data_redux}
Data reduction closely followed the methods of \citet{Milisavljevic13}. A series of scripts were written to automate IRAF/PyRAF\footnote{IRAF is distributed by the National Optical Astronomy Observatories, which are operated by the Association of Universities for Research in Astronomy, Inc., under cooperative agreement with the National Science Foundation. PyRAF is a product of the Space Telescope Science Institute, which is operated by AURA for NASA.} procedures to homogeneously reduce the spectral data. For each position, 2D images were trimmed, bias-subtracted, flattened, and co-added to remove cosmic rays. The L.A.Cosmic task \citep{Dokkum01} was employed to remove any residual cosmic rays. Images were then wavelength calibrated in the dispersion axis using comparison lamp images and straightened in the spatial axis using tracings of stellar continua.

Robust background subtraction was achieved with the IRAF task \texttt{background} using a sixth-order Chebyshev function fit along the spatial direction that was sampled from each individual pixel (i.e., no median pixel-binning). This resulted in sky emission-cleaned 2D images from which 1D spectra could then be extracted. An example of a fully-reduced and cleaned 2D image is shown in Figure \ref{fig:chip}.

\begin{figure*}[ht]
\centerline{%
\includegraphics[width=\linewidth]{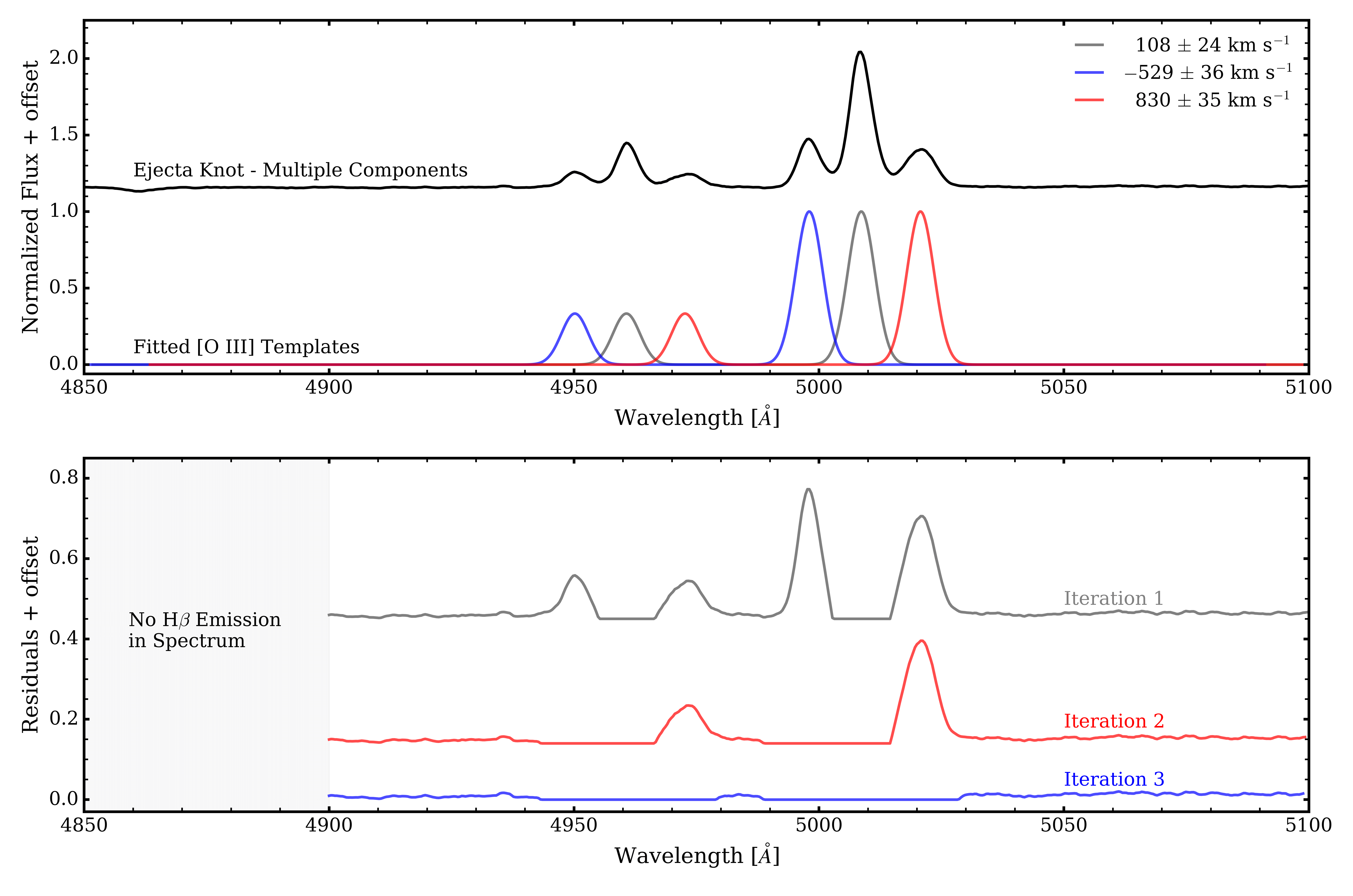}%
}%
\caption{\textit{Top}: An example of the cross-correlation method used to derive radial velocities of an O-rich ejecta knot with multiple velocity components. The ejecta spectrum has been smoothed, as described in the text, and is shown in black, while the three fitted [\ion{O}{3}] doublet templates are shown below in blue, gray, and red. \textit{Bottom}: Residuals for three iterations of the autocorrelation fitting process.} 
\label{fig:xcsao_example}
\end{figure*}

Each line of the 2D spectra was extracted in regions of conspicuous [\ion{O}{3}] $\lambda \lambda$4959, 5007 that were lacking in prominent H$\beta$ emission. This selection targeted O-rich ejecta. No attempt was made to catalog H$\beta$ emission associated with H-rich ISM/CSM. From these 1D extractions, knot velocities were measured through cross-correlation of templates using the IRAF task \texttt{xcsao}. The template was constructed by assuming a 3:1 intensity ratio of the [\ion{O}{3}] doublet \citep{Storey00}. We present example spectra of several types of ejecta knots encountered in Figure \ref{fig:example_spectra}. The \texttt{xcsao} task was run iteratively on all extracted 1D spectra for Doppler velocities ranging between $-$3000 and $+$3000 km s$^{-1}$ to identify individual knots. The spectra were first smoothed with a 7 by 7 boxcar before running the autocorrelation. This procedure was found to work well even in cases with multiple knots (e.g., Figure \ref{fig:example_spectra}, top right), and a detailed example of this analysis is shown in Figure \ref{fig:xcsao_example}. Measured velocities are believed to be accurate to $\pm 30$~km~s$^{-1}$ at the 68\% confidence level.

Results of the preliminary selection were manually inspected for false positives. It was important to only select emission that could be sensibly associated in 3D kinematic space. The Lasker's bowl region of the remnant and other modest-velocity ejecta knots ($\pm400$ km s$^{-1}$) required additional care owing to the significant presence of H$\beta$, and these data are considered separately than the more secure O-rich high-velocity knot detections. Regions where emission from O-rich knots were identified are shown in Figure \ref{fig:2d_emission_plot} and are labeled according to the notation from \citet{Morse95}. We refer the interested reader to \citet{Blair00} and \citet{Dopita18} for more detailed discussions of the spectroscopic properties of these types of knots.

A 8.3-magnitude earthquake struck approximately 500 miles south of Las Campanas Observatory on 16 September 2015 that disrupted optical alignment of GISMO within IMACS. There was no loss in data quality or coverage of the remnant, but the disruption did compromise use of pre-written pipeline procedures that stitch together the reformatted image slices. A variety of cross-checks ensured that consistent and accurate coordinates in right ascension (R.A.) and declination (Dec.) were assigned to radial velocity measurements for individual knots. Stars with well-measured coordinates encountered during the progression of long-slit positions provided fiducial reference points. Final fine-grain corrections to all positions were accomplished through comparison with high-resolution \textit{HST} images (Figure~\ref{fig:hstcomposite}) and overlays of bright stars from the 2MASS database \citep{Skrutskie06}, which provided accurate positional offsets. Positional uncertainties are estimated to be no more than $1.6^{\prime \prime}$, which is on the order of the slit width.

\section{Results} \label{sec:reconstruction}

\subsection{3D Doppler Reconstruction} \label{subsec:doppler_reconstruction}

\begin{figure*}[ht]
\centerline{%
\includegraphics[width=\textwidth]{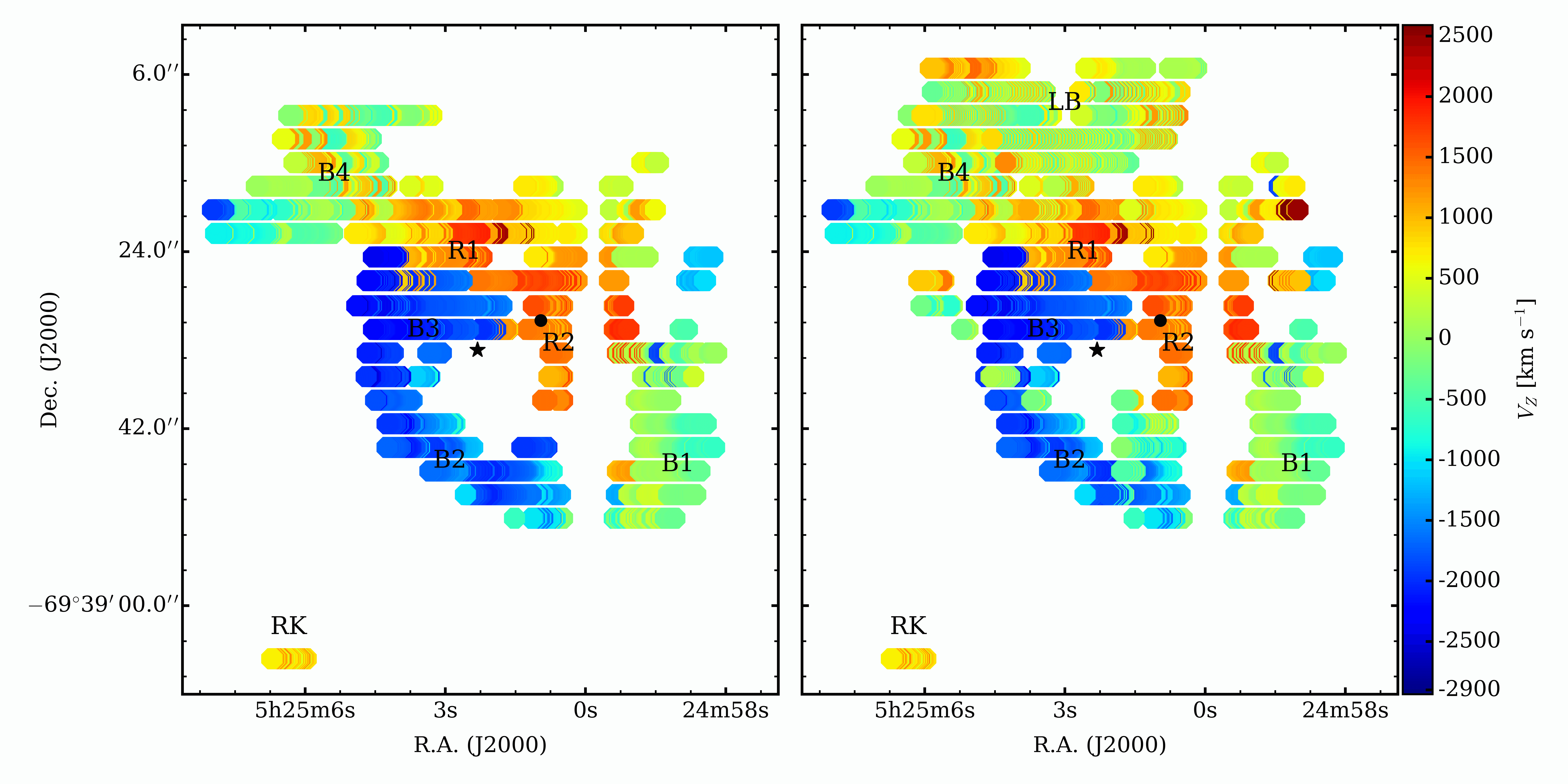}%
}%
\caption{Regions in N132D where O-rich knot emission was detected. Major knots from \citet{Morse95} have been identified along with the runaway knot (RK) discussed in the text. Both proposed centers of the remnant from \citet{Morse95} have been plotted, namely the center of the O-rich knots as a star ($\star$) and that determined by fitting the remnant's outer rim as a filled black dot ($\boldsymbol{\cdot}$). \textit{Left}: Only secure detections of pure O-rich knots are shown. \textit{Right}: Both secure detections as well as slow-velocity knots and those spatially associated with ISM/CSM-related clouds such as Lasker's Bowl (LB) \citep{Morse96} are shown. Pixels are not to scale and have been enlarged for visual clarity.}
\label{fig:2d_emission_plot}
\end{figure*}

A three-dimensional reconstruction of ejecta from a debris field requires knowledge of the center of expansion (COE). No proper motion analysis of N132D providing a COE from direct measurements of ejecta has been published. N132D has two suggested COEs inferred by different geometric assumptions: $\alpha(2000.0) = 5^h25^m02.^s7; \delta(2000.0)=-69^{\circ}38^{\prime} 34^{\prime \prime}$, which marks the center of the O-rich knots assuming a symmetric distribution, and $\alpha(2000.0)=5^h25^m01.^s4; \delta(2000.0)=-69^{\circ}38^{\prime}31^{\prime \prime}$, which corresponds to the remnant center determined from fitting an ellipse to the diffuse outer rim \citep{Morse95}. Figure~\ref{fig:2d_emission_plot} shows the location of these centers with respect to our observations. As this study is focused on O-rich knots, we assume N132D's explosion center to be located at the center of a symmetric distribution of O-rich ejecta. The two reported centers are separated by ${\sim}7^{\prime \prime}$, thus the choice between these COEs does not impact broader conclusions about the kinematics and morphology of N132D.

To simplify reconstruction, we also assume that the fast-moving O-rich ejecta follow  ballistic trajectories from the COE of the initial SN explosion. Even without proper motions, this is a reasonable assumption to make since individual O-rich knots are originally very dense compared to the surrounding ISM/CSM ($\chi > 100$) and the cloud shock velocities (${\sim}$100~km~s$^{-1}$) are much smaller than the typical O-rich ejecta velocities (${\sim}$1000--2000~km~s$^{-1}$). Such ballistic trajectories from the COE are described by the simple relation:

\begin{equation}
\label{eqn:1}
v = r / S,
\end{equation}

\noindent where $v$ is the radial velocity, $r$ is the angular distance from the explosion center, and $S$ is the scaling factor\footnote{We note that Equation \ref{eqn:1} is misprinted as multiplication of $S$ and $r$ in \citet{Milisavljevic13}, instead of division, as is correctly reported here.}. We adopt a distance of $d=50$~kpc to the LMC \citep{Panagia91, Marel02}, which implies a linear scale of 0.24~pc arcsecond$^{-1}$.

We determine the value of $S$ by fitting the measured Doppler velocities to a spherical expansion model. We closely follow the procedures reviewed in \citet{Milisavljevic13} and also discussed in \citet{Reed95} and \citet{DeLaney10}. In brief, we fit a semi-circle model to the velocity distribution and the observed projected radius. We parameterize the model as:

\begin{equation}
\label{eqn:2} 
\left( r_p / S \right)^2 + \left(v_D - v_c \right)^2 = \left( v_c - v_m \right)^2,
\end{equation}

where $v_c$ is the center of the velocity distribution, $v_m$ is the minimum velocity at which the semi-circle crosses the velocity axis, $r_p$ is the observed projected radius, $v_D$ is the observed Doppler velocity, and $S$ is a scaling factor relating the velocity axis to the spatial axis.

\begin{figure}[ht]
\centerline{%
\includegraphics[width=\linewidth]{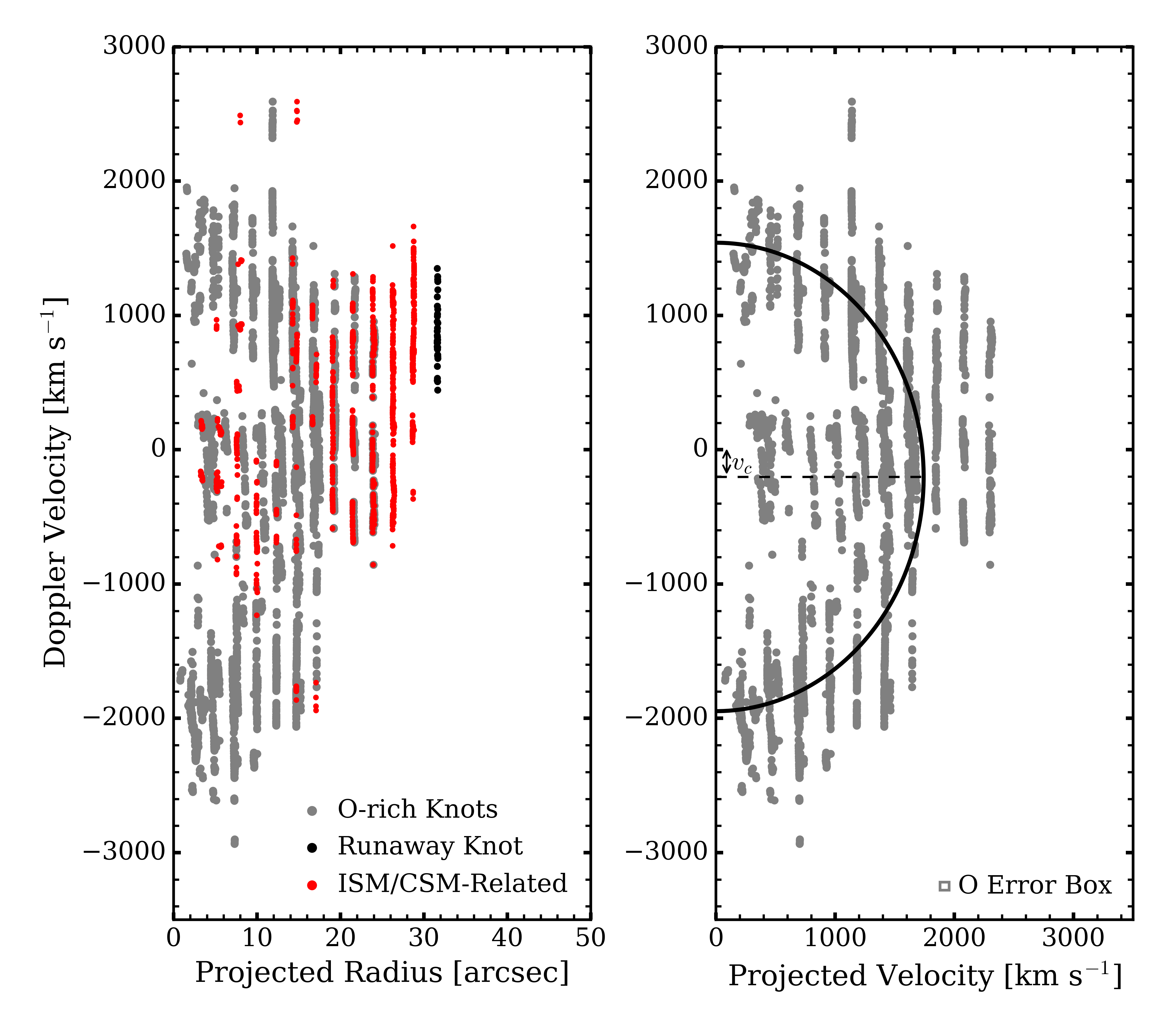}%
}%
\caption{\textit{Left}: Measured [\ion{O}{3}] ejecta velocities versus projected radii from the COE. O-rich ejecta knots are shown in gray, the runaway knot in black, and ISM/CSM-related material in red. \textit{Right}: Measured ejecta velocities versus projected radii, which have been converted to velocities using scaling factor $S=0^{\prime \prime}.010~\pm~0^{\prime \prime}.0005$ per km~s$^{-1}$. Best-fit semi-circle is shown as a solid black line.} 
\label{fig:lsq_fit}
\end{figure}

The results of our least-squares fit to the data are shown in Figure \ref{fig:lsq_fit}. The calculated Doppler velocities are $v_c=-203~\pm~20$~km~s$^{-1}$ and $v_m=-1947~\pm~150$~km~s$^{-1}$, and the scaling factor is $S=0^{\prime \prime}.010~\pm~0^{\prime \prime}.0005$ per km~s$^{-1}$. We fit the expansion model using only the secure detections of main shell O-rich knots, excluding the RK and lower-velocity material potentially associated with ISM/CSM-related clouds. Similarly, we restrict all further analysis to these secure detections; while we did not include the RK in the fitting process, we do include it in the subsequent velocity-space reconstructions. Knot positions in R.A.\ and Dec.\ were scaled to velocities using Equation \ref{eqn:1}. We used the COE and the calculated Doppler velocity $v_c$ to define a 3D COE in velocity space from which all reported vector trajectories originate.

\subsection{Age Estimation}
% D= $36.3^{\prime \prime}$
From our fitting in the previous subsection, the projected radius of N132D is $18.2^{\prime \prime}$, or 4.4~pc at the LMC distance of 50~kpc. In velocity space, this radius corresponds to a projected velocity of 1745~km~s$^{-1}$. Knowing the distance and the rate of expansion (assuming ballistic trajectories), we are able to estimate the age of the remnant using

\begin{equation}
\label{eqn:3} 
r_{\rm{vel}} [\rm{km}~\rm{s}^{-1}] \times t = r_{\rm{dist}} [\rm{pc}],
\end{equation}

\noindent where $r_{\rm{vel}}$ is the projected radius in km~s$^{-1}$, $t$ is time of expansion, and $r_{\rm{dist}}$ is the projected radius in pc. The time in terms of years is then given by:

\begin{equation}
\label{eqn:4}
t [\rm{yr}] =  9.785 \times 10^{5} \times \left( r_{\rm{dist}} [\rm{pc}] / r_{\rm{vel}} [\rm{km}~\rm{s}^{-1}] \right)
\end{equation}

\begin{figure*}[ht]
\centerline{%
\includegraphics[width=\linewidth]{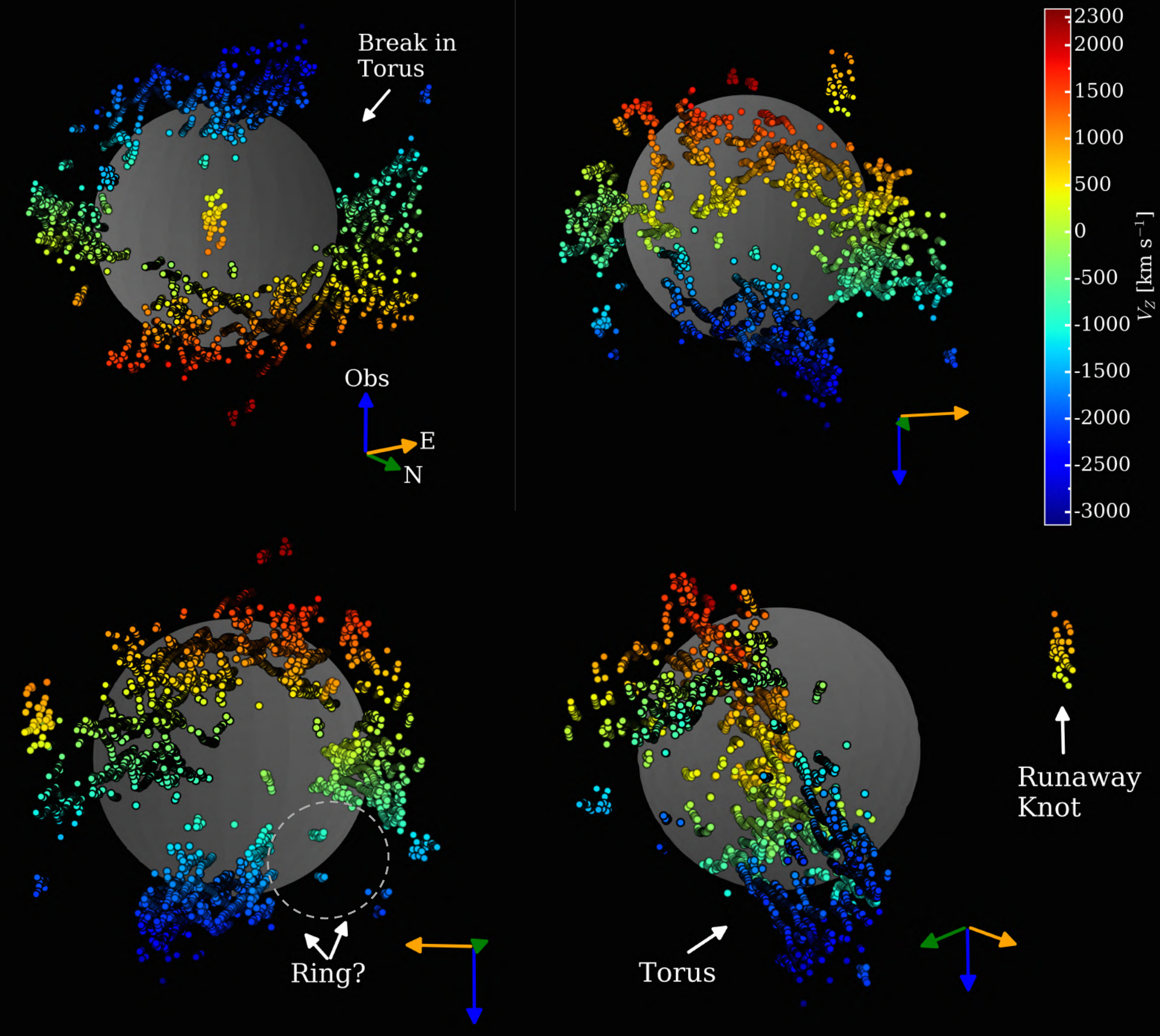}%
}%
\caption{3D Doppler reconstruction of N132D's optically-emitting, O-rich ejecta. The top left panel shows a clearer view of the broken torus structure. The top right and bottom left panels show two 45$^{\circ}$ views on each side of N132D. The bottom right panel presents a side profile of the torus, which highlights the orthogonal position of the runaway knot with respect to the plane of the tilted ring. A translucent sphere is a visual aid to help distinguish between front and back material.} 
\label{fig:N132D_views}
\end{figure*}

From these fitting results, we find an age of 2450\,$\pm\,195$ yr, which is consistent with the previously-determined kinematic age of 2500~yr from \citet{Vogt11}. Our age estimate is also consistent with recent X-ray and gamma ray observations indicating that N132D is undergoing the transition from a young to middle-aged SNR \citep{Bamba18}.

Compared to \citet{Vogt11}, our ring diameter is ${\sim}30\%$ percent smaller than their value of ${\approx}12$~pc, but our age estimates are consistent. The discrepancy in diameter is not unexpected, however, given that \citet{Vogt11} manually adjusted the expansion time in order to visually reproduce a ring morphology. \citet{Vogt11} reasoned that if the remnant is a true ring, its projected size in the X-Z plane should equal its major axis in the X-Y plane and was the criterion from which they derived their age estimate. Our method calculates the ring diameter using an average of all points measured from data having higher spatial resolution and improves the overall fitting procedure. If we instead only fit the outer limb of the tilted ring in our data set, we also find a diameter of ${\approx}48^{\prime \prime}$ ($12$~pc). Alternatively, if we fit only the inner limb of this ring, we would find a diameter of ${\approx}24^{\prime \prime}$ ($6$~pc).

We found a conspicuous lack of O-rich ejecta with radial velocities between $-$1000 to $-$2000~km~s$^{-1}$ and projected velocities of ${\sim}$2000~km~s$^{-1}$ (Figure \ref{fig:lsq_fit}). We investigated whether there were any significant gaps in our coverage at or near this phase space that would adversely affect our least squares fit of the ring diameter. We confirmed that there were no such gaps in our coverage. Instead, the absence of such data points is either the result of faint [\ion{O}{3}] knots that were not detected at the sensitivity of our survey, or is related to intrinsic inhomogeneities in the spatial distribution of O-rich ejecta. We favor this latter explanation, especially since ejecta with radial velocities in this range are found to be associated with an incomplete ring-like structure toward the blue-shifted side of N132D (see Section \ref{sec:kinematic_str}.)

\subsection{O-rich Ejecta}  \label{sec:results}

The results from our survey of N132D encompass 4126 individual data points with 4093 corresponding to main shell ejecta and 33 for the RK. Measured knot radial velocities are shown in Figure \ref{fig:N132D_views} with a color-coded gradient. The resulting Doppler maps represent the most complete catalog of its [\ion{O}{3}] optically-emitting ejecta to date. 

\begin{figure*}[ht]
\centerline{%
\includegraphics[width=\linewidth]{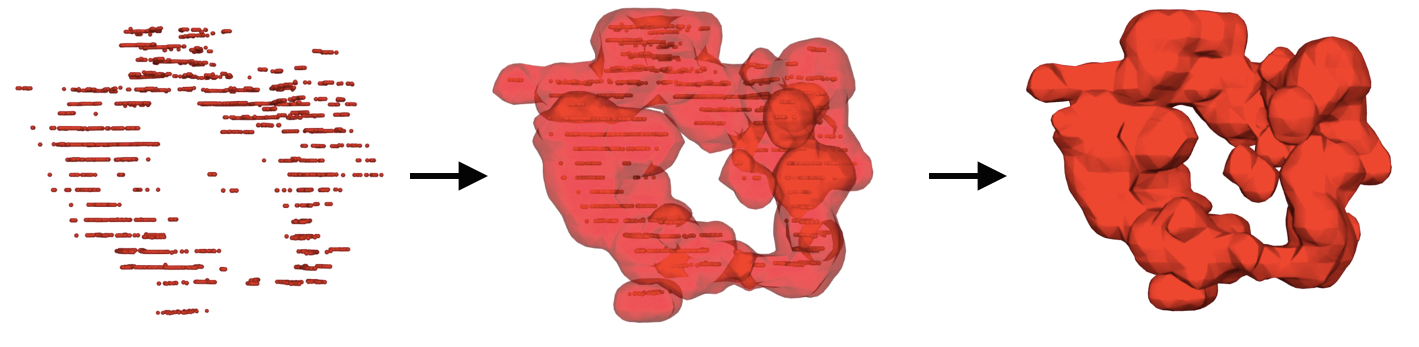}%
}%
\caption{Surface reconstruction process for the oxygen-emitting ejecta in N132D. The three steps shown illustrate the process from initial point cloud measured spectroscopically to the surface model derived using a Marching Cubes algorithm and then retopologization with Catmull-Clark smoothing. An animation of this surface reconstruction can be found in the supporting materials of this paper.} 
\label{fig:reconstruction_anim}
\end{figure*}

An animation of the radial velocity point cloud that has been enhanced with a surface model is provided in the supporting material for this paper. The surface model is first created using the Marching Cubes method and further retopologized with Catmull-Clark smoothing. The Marching Cubes algorithm uses an input threshold to polygonize a three-dimensional scalar field using a table of predefined facet configurations \citep{Lorensen87}. Catmull-Clark operates by subdividing the surface of a mesh into smaller polygons and readjusting the vertices upon weighted averages \citep{Catmull98}. By performing such a surface reconstruction, we aim to produce a more faithful reconstruction of the gas distribution, and thus a more physically accurate 3D representation of N132D. A demonstration of this surface reconstruction process is shown in Figure \ref{fig:reconstruction_anim}. 

Below, we discuss some specific characteristics of N132D that are revealed by our high-resolution 3D reconstruction of the remnant's kinematic structure.

\subsubsection{Kinematic Structure} \label{sec:kinematic_str}

The majority of O-rich knots define a broken and distorted torus. The large-scale morphology is consistent with the 3D models presented in \citet{Vogt11} based on [\ion{O}{3}] $\lambda5007$ emission. The main difference between our 3D reconstruction and those of \citet{Vogt11} is increased spatial resolution by a factor of ${\approx}$2--3 with comparable kinematic resolution, and deeper exposures sensitive to fainter emission and finer structure. Our study also leverages the emission from both lines of the [\ion{O}{3}] doublet, allowing for more secure detections of true O-rich knots and an increased ability to more easily disentangle complex, multi-velocity component knots.

\begin{figure*}[ht]
\centering
\includegraphics[scale=0.0575]{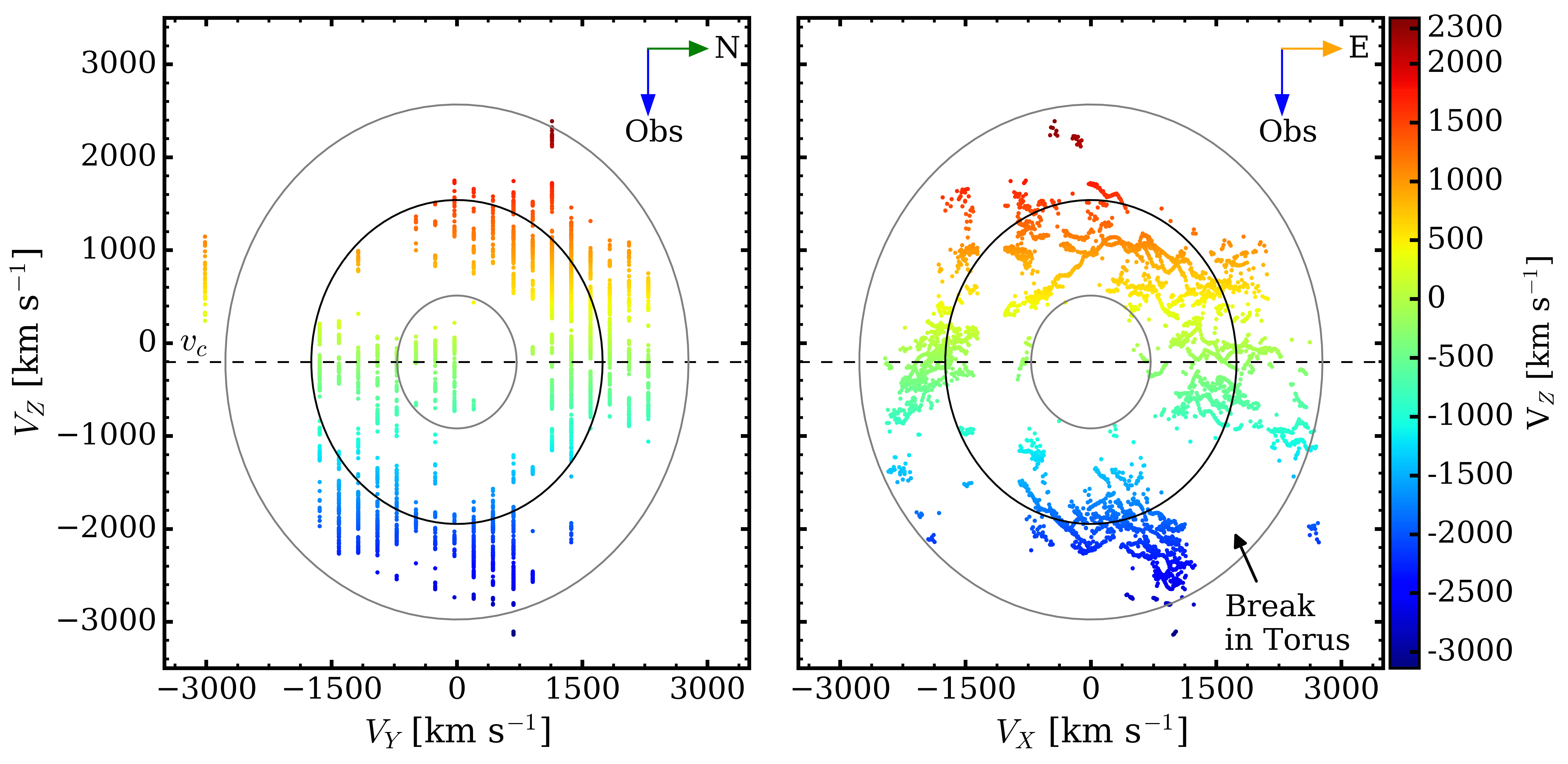}
\includegraphics[scale=0.042]{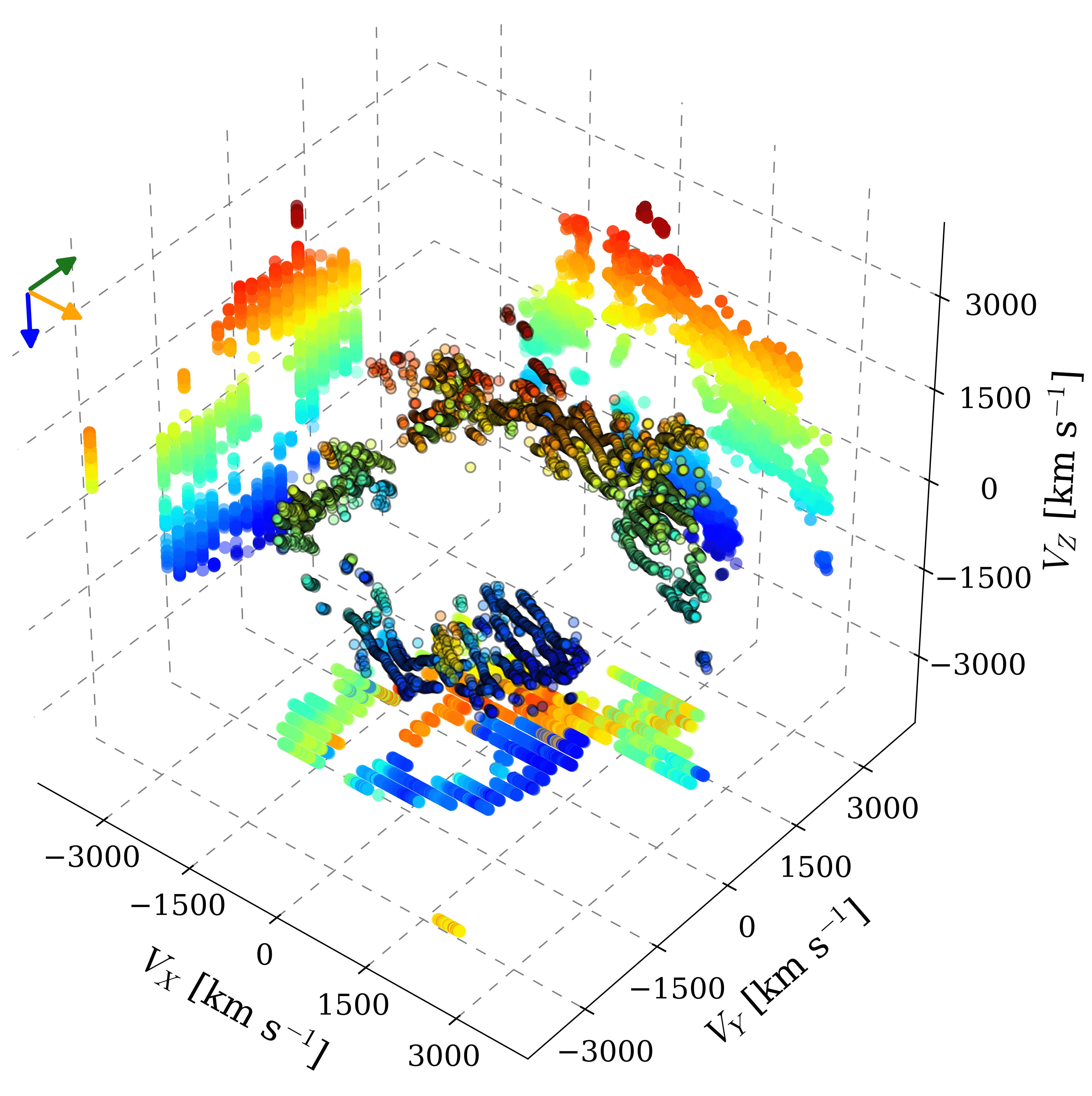}
\caption{Projections in the $V_Y$-$V_Z$ (\textit{left}) and $V_X$-$V_Z$ planes (\textit{middle}). The best-fit torus is overlaid with major and minor radius shown in black and gray, respectively.  Three-dimensional velocity space plot with projections in all three planes (\textit{right}).} 
\label{fig:plane_views}
\end{figure*}

Figure \ref{fig:plane_views} shows various projections of N132D, which illustrate the remnant's broken torus structure. The bulk ejecta of N132D exhibit a blue-shifted radial velocity asymmetry of $-$3000 to $+$2300~km~s$^{-1}$, consistent with previous kinematic studies \citep[e.g.,][]{Lasker80, Sutherland95}. This velocity asymmetry of 700~km~s$^{-1}$ is nearly twice that determined by \citet{Vogt11}, and is likely due to our more sensitive and spatially-complete survey detecting additional high-velocity knots.

\subsubsection{Best-Fitting Torus}

To quantitatively characterize the toroidal structure, we fit a torus model to the kinematic reconstruction of N132D. We fix the major radius to the best-fit semi-circle radius of 1745~km~s$^{-1}$, as determined in Section \ref{subsec:doppler_reconstruction}. In order to constrain the inclination of the torus and its minor radius, we employ a Monte Carlo (MC) approach to find the minimum volume torus that contains the maximum number of main shell points, excluding emission associated with the RK. First, for a fixed minor radius, we randomly select a set of inclination angles, $\theta_x, \theta_y, \theta_z$, each between 0 and 360$^{\circ}$ and calculate the percentage of points contained within the best-fitting torus. This process is then repeated for a large number (${\sim}10^5$) of individual trials. We then choose the best-fitting torus inclination as the set of inclination angles that maximizes the percentage of points contained within the torus. Once a torus inclination is determined, we vary the value of the minor radius between 400 and 1500~km~s$^{-1}$ in steps of 10~km~s$^{-1}$ and for each value of minor radius, we again calculate the percentage of points contained within the best-fitting torus. 

\begin{figure*}[ht]
\centerline{%
\includegraphics[width=\linewidth]{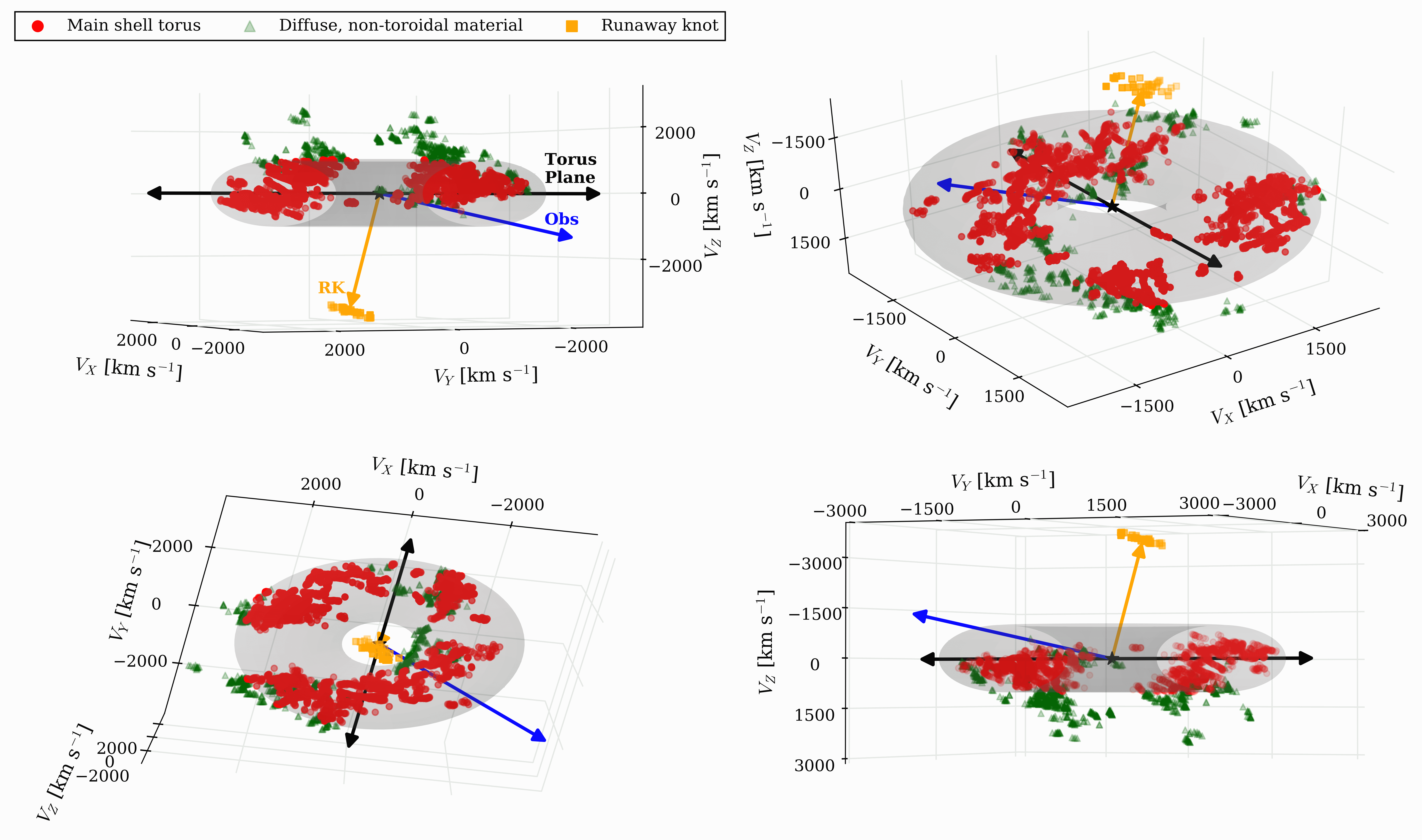}%
}%
\caption{Various perspectives of the best-fitting torus. The top left panel shows an edge-on view, while the top right panel shows a 45$^{\circ}$ view. A top-down view is shown in the bottom left and the bottom right panel is another edge-on view. Main-shell ejecta that are well-fitted by a torus are labeled in red and diffuse, scattered material that do not strictly follow a torus distribution are shown in green. The runaway knot is shown in orange.} 
\label{fig:torus_fitting}
\end{figure*}

We define the minor radius of the torus such that it contains 75\% of the main shell ejecta points by interpolating a minor radius that corresponds to 75\% of enclosed points. This approach disfavors scattered material that does not strictly follow the torus distribution (highlighted in green in Figure \ref{fig:torus_fitting}).  To estimate uncertainties, we repeat this calculation using an MC method, where the uncertainty of each 3D velocity point is modeled as a Gaussian with a width correspond to its velocity error and then for each point, a value is randomly chosen. The final value and uncertainties are reported as the median and upper and lower quartile values. We find a minor radius of 1028$^{+54}_{-25}$~km~s$^{-1}$, which corresponds to a total width of ${\sim}$2060~km~s$^{-1}$, or 4.9~pc in physical extent. With respect to a unit normal vector in the plane of the sky oriented from the remnant to the observer, the normal vector from the torus' equatorial plane is inclined $28^{\circ} \pm 5^{\circ}$ to the line of sight. Our measurement of the inclination angle is consistent with that of \citet{Vogt11}, who estimated ${\sim}25^{\circ}$ via visual inspection.

There are notable deviations from a simple torus model. Perhaps most prominent is that the vertical structure of N132D varies with azimuthal angle. Specifically, the redshifted material is elevated above the mid-plane of the torus, while the blue-shifted side is below the mid-plane. We discuss these deviations further in Section \ref{sec:discussion}.

We loaded the data into our Virtual Reality Laboratory (VR Lab; McGraw et al., in preparation) and explored the 3D morphology of N132D for evidence of ejecta substructure in the form of rings or bubbles. For instance, multi-wavelength reconstructions of Cas~A have revealed large-scale coherent rings and bubbles of ejecta substructure, indicative of a bubble-like interior powered by radioactive $^{56}$Ni-rich ejecta \citep{DeLaney10, Milisavljevic13, MF15, Wongwathanarat17}. Other SNRs, including E0102 and 3C58, also exhibit ring-like structures to varying degrees \citep{Finkelstein06,Milisavljevic16-IAU,LF18}. We were unable to conclusively identify any such substructure in N132D. However, an incomplete ring is seen toward the blue-shifted side of N132D and is composed of diffuse O-rich material that deviates from the larger torus-shaped morphology. This feature has a radius of 1000~km~s$^{-1}$, which is approximately 60\% of the torus radius and is labeled in the lower left panel of Figure \ref{fig:N132D_views}.

\subsubsection{Runaway Knot}

With respect to a unit normal vector from the torus' equatorial plane, the RK is inclined by approximately $82^{\circ} \pm 2^{\circ}$. The RK has a median Doppler velocity of ${\sim}$820~km~s$^{-1}$ and a high median total space velocity of 3650~km~s$^{-1}$. This is approximately twice the velocity of the bulk ejecta traveling at 1745~km~s$^{-1}$.

\begin{figure}[ht]
\centerline{%
\includegraphics[width=0.55\textwidth]{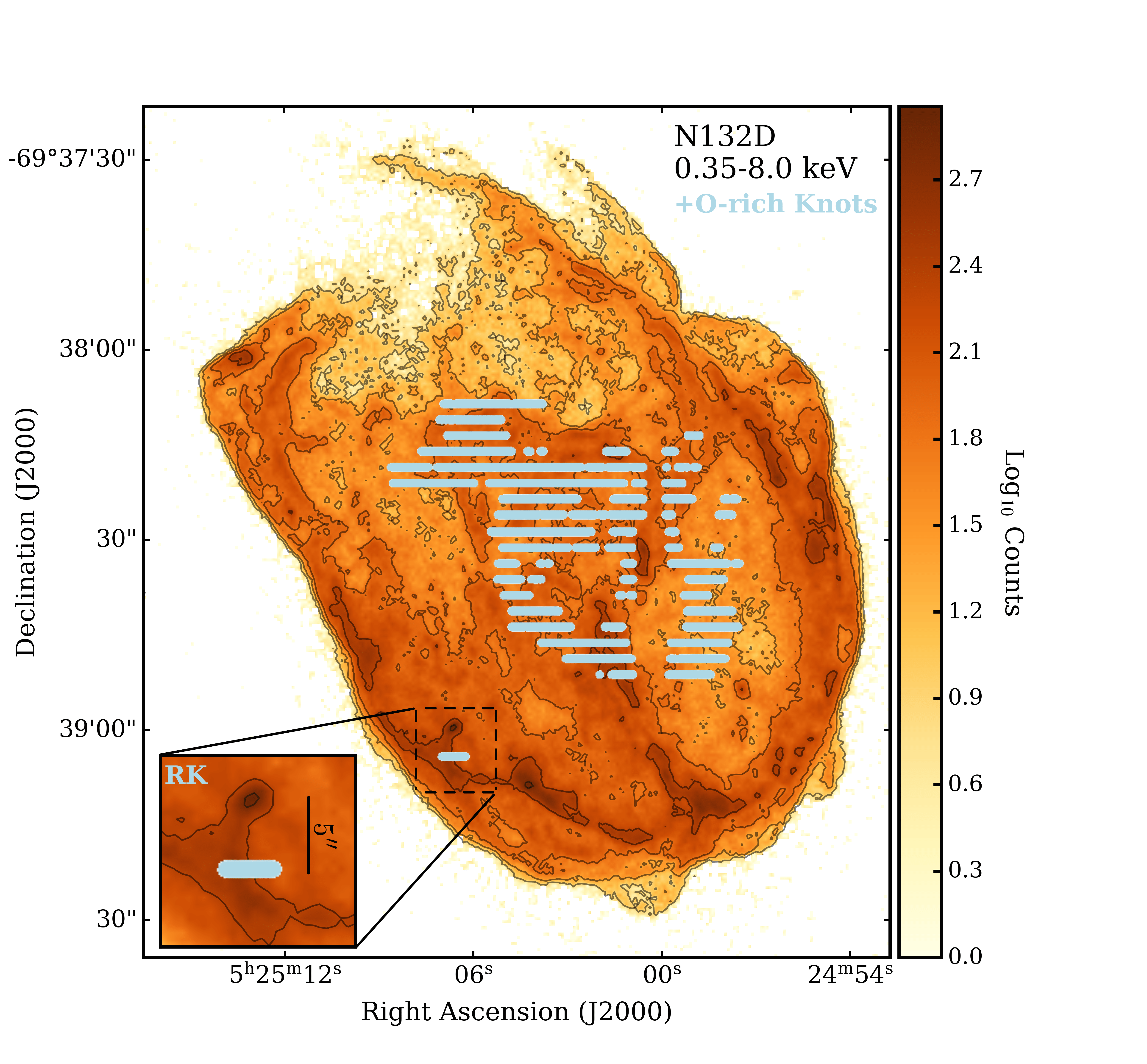}%
}%
\caption{\textit{Chandra} Advanced CCD Imaging Spectrometer S-array (ACIS-S) image of counts per pixel in N132D in the 0.35--8.0~keV band with O-rich optical ejecta overlaid. Each pixel is $0.5^{\prime \prime}$. Inset shows the $5^{\prime \prime}$ (${\sim}1.2$~pc) offset between point source-like X-ray emission and the RK.}
\label{fig:hotspot}
\end{figure}

The RK is also in close proximity with a point source-like X-ray enhancement in the south-east of the remnant. Figure \ref{fig:hotspot} shows a \textit{Chandra} image in the 0.35--8.0~keV band from 89~ks of archival data \citep{Borkowski07} with our O-rich ejecta overlaid. This X-ray enhancement is located only ${\sim}5^{\prime \prime}$ from the RK and is also close to a bright filament of X-ray emission. Offsets between optical and X-ray emission in SNRs are not uncommon \citep{Patnaude14} and are associated with differences in temperature and density, as optical and X-ray emission arises from plasmas with different physical conditions. The exact relationship between the X-ray bright knot and RK (as well as the bright X-ray filament) is not clear, but our data suggest a physical association.

This X-ray enhancement is the brightest knot observed in N132D in the 0.35--8.0~keV band, while our optical spectrum of the RK shows only weak [\ion{O}{3}] $\lambda\lambda$4959, 5007 emission. A spectral analysis of the \textit{Chandra} data indicates that this knot is enhanced in Si ($0.80_{-0.08}^{+0.14}$) and slightly enhanced in S ($0.55_{-0.14}^{+0.14}$), but has a lower O abundance ($0.10_{-0.10}^{+0.33}$). This contrasts with regions near the forward shock that are well-fitted with abundances of $0.4\times$ Solar, which is typical of the ISM in the LMC. However, a definitive assessment of both the chemical abundances of this X-ray enhancement and its relationship with the RK will require X-ray data with at least an order of magnitude improvement in spectral resolution and comparable or better spatial resolution than the existing \textit{Chandra} data.

\section{Discussion} \label{sec:discussion}

Linking the kinematic and chemical characteristics of SNRs to the explosion processes associated with their parent SNe is rapidly growing in interest \citep{Gabler17,MF17,Wongwathanarat17,Ferrand19}. However, there are many challenges to understanding the manifold morphologies of SNRs that reflect the diversity of their progenitor star systems, prior mass loss, and environmental conditions (including, e.g., density and metallicity) that can lead to conflicting interpretations (see, e.g., \citealt{Lopez13} vs.\ \citealt{Zhou18}; \citealt{Broersen14} vs.\ \citealt{Gvaramadze17}). Typically, distributions of elements at X-ray wavelengths have been used as a classification diagnostic, and as a way to make inferences about explosion energies, nucleosynthesis yields, and density of the surrounding material \citep{LF18}. Recent advances of this strategy include use of Fe-K line luminosities and energy centroids \citep{Yamaguchi14,Patnaude15} and mapping the distribution of ejecta with respect to remnant compact objects \citep{Fesen06,Katsuda18,Holland-Ashford19}. Light echo spectroscopy has made it possible to directly connect SNRs with their parent SN classifications \citep{Krause08,Krause08-Tycho,Rest08} and to observe the explosion from multiple lines of sight \citep{Rest11}. 

Most investigations of SNRs characterize them as two-dimensional projections observed on the sky.  However, dense spectroscopic mapping at optical \citep{Vogt10,Milisavljevic13}, near-infrared \citep{Kjaer10,DeLaney10,MF15,Larsson16}, X-ray \citep{DeLaney10,Grefenstette17}, and radio wavelengths \citep{Abellan17} has made it possible to develop 3D reconstructions of SN ejecta morphology. These developments have motivated substantial efforts to link explosions to their remnants with end-to-end hydrodynamic / magnetohydrodynamic models \citep{Orlando16,Orlando19}. Our kinematic reconstruction of N132D contributes testable constraints for such simulations. 

Many simulations have focused on explosion asymmetry introduced at the time of core-collapse by neutrinos, fluid instabilities, rotation, and
magnetic fields into  spherically symmetric progenitor stars and environments \citep{Woosley05,Janka12}. However, observations and theoretical predictions have motivated additional considerations \citep{Margutti14,Couch14,Nakamura15,Summa16,McDowell18}. It now seems likely that progenitor structure is significantly perturbed by currently poorly understood terminal phases of stellar evolution \citep{Arnett11,SA14}, and revisions to our understanding of mass loss and observations of a high binary fraction among O-type stars \citep{Smith14} suggest circumstellar environments that are not spherically symmetric \citep{SP96,Ryder04}. Below we discuss these issues in the context of the observed morphology of N132D and attempt to discern whether the large-scale structure is reflective of the original explosion kinematics and dispersal of elements, or instead has been more strongly influenced by interaction with mass loss of the progenitor star prior to core collapse.

\subsection{Geometry of O-rich Knots and Large-scale Velocity Asymmetry}

The dominant large-scale coherent structure of N132D is a distorted torus with a broken circumference. Cas~A has long been known to have a general torus-like distribution of ejecta \citep{Markert83,DeLaney10} with ring- and bubble-like substructure \citep{Lawrence95,Milisavljevic13,MF15}. It is unclear whether this torus distribution is intrinsic to the explosion or is the result of interaction with an inhomogenous environment. \citet{DeLaney10} argue that the ejecta morphology is the result of a flattened explosion where the highest velocity ejecta were expelled in a thick torus tilted not far off the plane of the sky in a number of large-scale pistons. However, high resolution mapping of Cas~A at infrared wavelengths show deviations from a strict torus distribution \citep{Isensee10}, questioning this interpretation \citep{Milisavljevic13}.  Later, \citet{MF15} mapped photoionized inner ejecta, as revealed in [\ion{S}{3}], that were also inconsistent with a flattened explosion.  

\begin{figure*}[ht]
\centerline{%
\includegraphics[width=\linewidth]{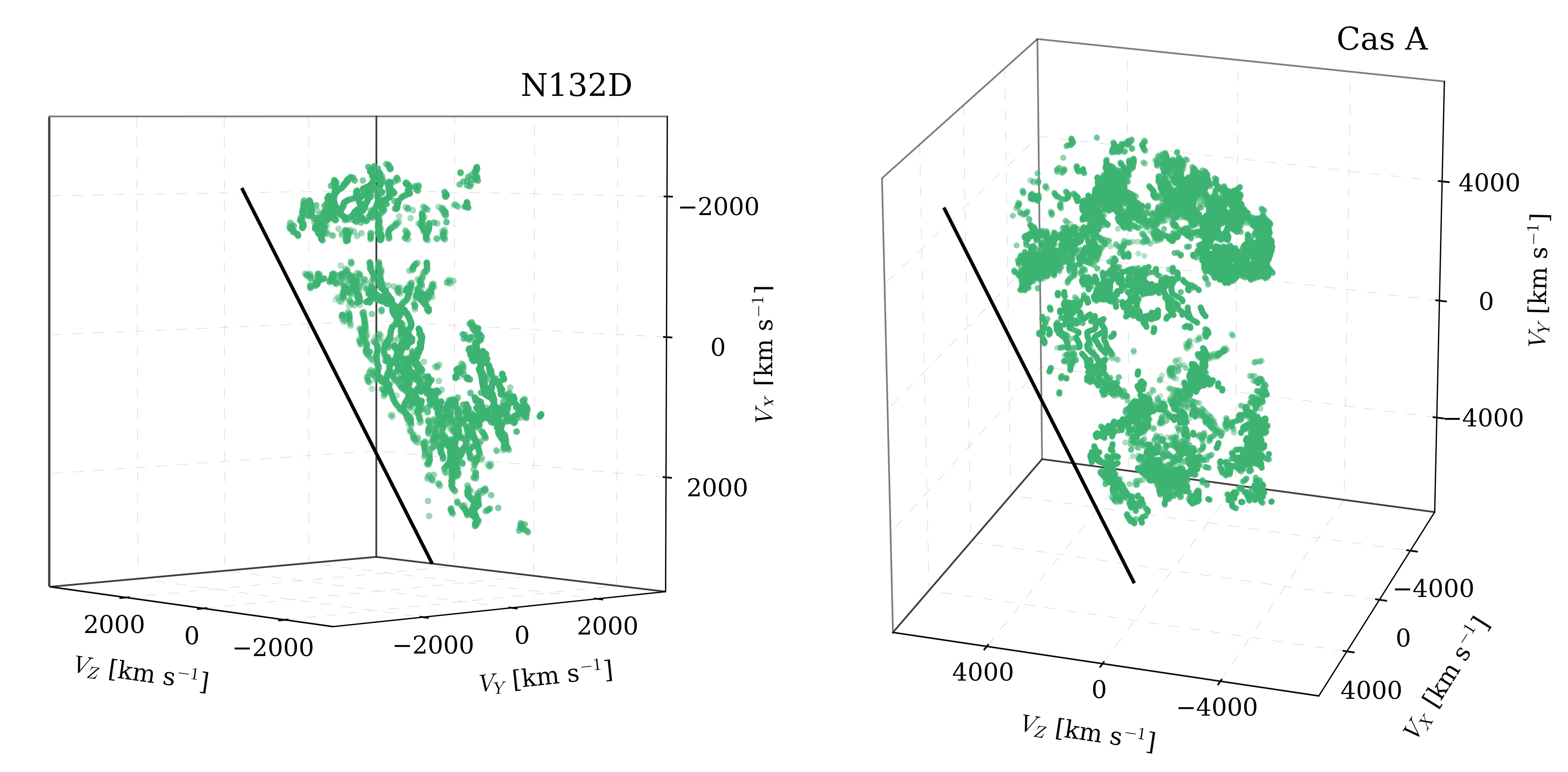}%
}%
\caption{N132D (\textit{left}) and Cas~A (\textit{right}) rotated at angles to illustrate their similar flat-sided torus distributions of main shell ejecta. The black lines indicate an abrupt cutoff in ejecta at a boundary defined by an angled plane. Cas~A data are taken from \citet{Milisavljevic13}.} 
\label{fig:Cas_A_boundary_compare}
\end{figure*}

The break in the torus of N132D is not unlike discontinuities observed in other remnants. In Cas~A, a conspicuous rupture in the northeastern portion of main shell ejecta located at the base of high-velocity material has persisted over decades of observations \citep{BM54,Patnaude14}. A similar rupture-like region is observed in E0102, but is not known to be associated high-velocity ejecta \citep{Finkelstein06}.  These two remnants exhibit additional substructure in the way of rings and bubbles that N132D does not \citep{Eriksen01,MF15}, which may be attributable to N132D's advanced age compared to Cas~A (age $\approx 340$ yr; \citealt{Fesen06}) and E0102 (age $\approx 1500-2000$ yr; \citealt{Finkelstein06}).

The Doppler velocity asymmetry of N132D is not uncommon in SNRs. Cas~A has a $-$4000 to $+$6000 \kms\ radial velocity asymmetry in its optical ejecta \citep{Milisavljevic13}, and E0102 has a velocity asymmetry spanning $-$2500 to $+$3500 \kms \citep{Vogt10}. \citet{Reed95} argued that the velocity asymmetry in Cas~A was due to an inhomogeneous environment.  If the density of the surrounding medium is greater on the blueshifted near side, forward expansion is inhibited and results in an apparently redshifted COE.  However, \citet{DeLaney10} and \citet{Isensee10} concluded that structural differences between the front and back interior surfaces could best be understood as having originated in asymmetries present in the explosion. In support of this view, the observed distribution of radioactive $^{44}$Ti measured by NuSTAR, which should not be strongly influenced by the SN interaction with its environment, has a bulk line-of-sight Doppler velocity of 1100--3000 \kms \citep{Grefenstette14}.  

An abrupt cutoff in the ejecta along a plane in the redshifted side of N132D was discovered while exploring the remnant in the VR Lab. The distribution is not unlike that observed in Cas~A. In Figure \ref{fig:Cas_A_boundary_compare}, we show N132D and Cas~A rotated at angles that highlight this shared morphology. \citet{Milisavljevic13} reported that the rear-facing ejecta of Cas~A cutoff abruptly along an angled plane, while the front-facing ejecta do not. This sharp cutoff may be most easily understood as the result of interaction with an inhomogeneous environment. However, rotation and interaction with a binary companion can also produce non-spherical mass loss scenarios \citep{Langer12}.

\subsection{High-Velocity Material}

Our survey, which covers the entire ${\sim}3^{\prime} \times 3^{\prime}$ field of N132D, was unable to uncover any high-velocity material in addition to the previously discovered RK. Given that (1) our survey missed 33\% of the remnant (Section \ref{sec:obs}), (2) was only sensitive to [\ion{O}{3}] $\lambda\lambda$4959, 5007 line emission, and (3) the RK is a relatively faint region of gas, it remains unclear if the RK is indeed an isolated knot or if a greater population of high-velocity material is present. The RK's close proximity to localized X-ray emission enriched in Si and S suggests that  emission lines associated with these elements may be better suited for detection. 

The RK may have a similar origin to the high-velocity ejecta in Cas A, which travel at speeds of upwards of 15,000 \kms\ (approximately three times faster than the main shell ejecta) and are enriched in S, Si, Ar, and Ca emission \citep{FG96}. Although their broadness and kinetic energy argue against the Cas A SN being a jet-induced explosion \citep{Laming06,FM16}, the jets are kinematically and chemically distinct from the rest of the remnant \citep{Fesen01,Hwang04}. This may reflect an origin in a jet-like mechanism that accelerated interior material from a Si-, S-, Ar-, and Ca-rich region near the progenitor's core up through the mantle and H-, He-, N-, and O-rich outer layers with velocities that greatly exceeded that of the rapidly expanding photosphere. 

The RK is also not unlike ejecta clumps or ``bullets'' located in the northeastern edge of the Vela remnant, which exhibit a high Si abundance, in contrast to the other observed ejecta fragments that show enhanced O, Ne, and Mg abundances \citep{Aschenbach95,KT06}. Recently, another ejecta fragment located opposite to shrapnel A with respect to the center of the shell was found in the southwestern boundary of the remnant, strongly suggesting a Si-rich jet-counterjet structure \citep{Garcia17}.

Our findings are possibly relevant to the continuum of explosion energies, extending from broad-lined Type Ic SNe associated with gamma-ray bursts to more ordinary Type Ib/c SNe, that appears to exist \citep{Mazzali08,Margutti14,Milisavljevic15,Sobacchi17,Barnes18} and that may be associated with energy input from compact objects formed
during core-collapse. The connection may also extend to superluminous SNe \citep{Milisavljevic13-12au,Mazzali14,Metzger15,Nicholl16,Milisavljevic18}. The implication is that a wide variety of
jet activity may potentially be occurring at energies that participate in but do not necessarily drive explosions \citep{FM16,Ertl19}. At extragalactic distances, explosions with weak or choked jets that lack sizeable relativistic ejecta are indistinguishable from events without jets \citep{Lazzati12}. Only through resolved inspection of nearby supernova remnants
can the presence of these structures be identified. Origins in the diversity of jet activity may be related to the type of compact object formed (neutron star vs.\ magnetar vs.\ black hole) and whether the central engine activity becomes choked before the jet is able to pierce through the stellar envelope.

\section{Conclusions}
\label{sec:conclusions}

We have presented a 3D kinematic reconstruction of optical emission from oxygen-rich ejecta in SNR N132D based on radial velocity measurements extracted from long-slit spectra. Based on this high spatial and kinematic resolution data set, we conclude the following:

\begin{enumerate}
\item We confirm the findings of previous kinematic studies of N132D that show the bulk of the remnant's optically bright oxygen-rich ejecta are arranged in a torus-like geometry. We find that this torus-shaped material is tilted by 28$^{\circ}$ with respect to the plane of the sky, has a radius of 4.4~pc ($D_{\rm{LMC}}$/50~kpc), exhibits a blue-shifted radial velocity asymmetry of $-$3000 to $+$2300~km~s$^{-1}$, and has a conspicuous break in its circumference.
\item Assuming homologous expansion from the geometric center of O-rich filaments, the average expansion velocity of 1745~km~s$^{-1}$ translates to an age since explosion of 2450~$\pm$~195~yr, which is consistent with the kinematic age determined from the lower spatial resolution study by \citet{Vogt11}.
\item A faint, spatially-separated ``runaway knot" with total space velocity of 3650~km~s$^{-1}$ is nearly perpendicular to the plane of N132D's main torus and coincident with localized X-ray emission that is enhanced in Si relative to the LMC and N132D's main ejecta. The kinematic and chemical signatures suggest that the RK may have had its origin deep within the progenitor star. 
\item N132D has notable kinematic and chemical similarities with Cas~A. Both have a  general torus morphology with a rupture, a sharp cutoff at a boundary defined by an angled plane, and high-velocity Si-enriched knots. 
\end{enumerate}

Our results underscore the need for SN-to-SNR simulations that take into consideration evolutionary phases and the associated mass loss that massive stars experience approaching core collapse. Our analyses strongly suggest that environment is an important contribution to the observed morphology. Further work comparing and contrasting morphology and chemical abundances of additional SNRs will provide valuable constraints for simulations attempting to model and correctly interpret the electromagnetic+gravitational wave+neutrino signals of SNe.

\acknowledgments

The authors thank the anonymous referee for valuable comments that improved both the content and presentation of this work. This work has benefited from the helpful comments of Fr\'{e}d\'{e}ric Vogt and Robert Fesen. This paper includes data gathered with the 6.5 meter Magellan Telescopes located at Las Campanas Observatory, Chile. STSDAS and PyRAF are products of the Space Telescope Science Institute, which is operated by AURA for NASA. This publication makes use of data products from the Two Micron All Sky Survey, which is a joint project of the University of Massachusetts and the Infrared Processing and Analysis Center/California Institute of Technology, funded by the National Aeronautics and Space Administration and the National Science Foundation. This work is also based on observations made with the NASA/ESA \textit{Hubble Space Telescope}, obtained from the Data Archive at the Space Telescope Science Institute, which is operated by the Association of Universities for Research in Astronomy, Inc., under NASA contract NAS 5-26555. These observations are associated with program 12001. The work also incorporates data obtained from the \textit{Chandra} Data Archive and observations made by the \textit{Chandra} X-ray Observatory. C.J.L. acknowledges funding from the National Science Foundation Graduate Research Fellowship under Grant DGE1745303.


\vspace{5mm}
\facilities{Magellan:Baade(IMACS$+$GISMO), CXO}


\software{DS9 \citep{Joye03}, IRAF/PyRAF \citep{Tody86}, L.A.Cosmic \citep{Dokkum01}, \texttt{Matplotlib} \citep{Hunter07}, \texttt{NumPy} \citep{van_der_Walt11}}

\bibliography{N132D.bib} % if your bibtex file is called example.bib



\end{document}